\documentclass[oldversion]{aa}
\usepackage{txfonts}
\usepackage{graphicx}
\usepackage{natbib}
\bibpunct{(}{)}{;}{a}{}{,}
\begin{document}

\title{Mode identification from monochromatic amplitude and phase variations for the rapidly pulsating subdwarf B star \\ EC 20338$-$1925\thanks{Based on observations collected at the European Organisation for Astronomical Research in the Southern Hemisphere, Chile (proposal ID 083.D-0415).}
}
\author{
S.K. Randall \inst{1}
\and G. Fontaine \inst{2}
\and P. Brassard \inst{2}
\and V. Van Grootel \inst{3}
}

\institute{
ESO, Karl-Schwarzschild-Str. 2, 85748 Garching bei M\"unchen, Germany; \email{srandall@eso.org}
\and D\'epartement de Physique, Universit\'e de Montr\'eal, C.P. 6128, Succ. Centre-Ville, Montr\'eal, QC H3C 3J7, Canada
\and Laboratoire d'Astrophysique de Toulouse-Tarbes, Universit\'e de Toulouse, CNRS, 14 Avenue \'Edouard Belin, 31400 Toulouse, France
}
\date{Received date / Accepted date}

\abstract
{
We obtain time-series spectrophotometry observations at the VLT with the aim of partially identifying the dominant oscillation modes in the rapidly pulsating subdwarf B star EC 20338$-$1925 on the basis of monochromatic amplitude and phase variations. 

From the data gathered, we detect four previously known pulsations with periods near 147, 168, 126 and 140~s and amplitudes between 0.2 and 2.3 \% of the star's mean brightness. We also determine the atmospheric parameters of EC 20338$-$1925 by fitting our non-LTE model atmospheres to an averaged combined spectrum. The inferred parameters are $T_{\rm eff}$ = 34,153$\pm$94~K, $\log{g}$=5.966$\pm$0.017 and $\log{N(\rm He)/N(\rm H)}$ = $-$1.642$\pm$0.022, where the uncertainty estimates quoted refer to the formal fitting errors. Finally, we calculate the observed monochromatic amplitudes and phases for the periodicities extracted using least-squares fitting to the light curves obtained for each wavelength bin. 

These observed quantities are then compared to the corresponding theoretical values computed on the basis of dedicated model atmosphere codes and also taking into account non-adiabatic effects. We find that the quality of the data is sufficient to identify the dominant pulsation at 146.9~s as a radial mode, while two of the lower amplitude periodicities must be low-degree modes with $\ell$=0-2. This is the first time that monochromatic amplitudes and phases have been used for mode identification in a subdwarf B star, and the results are highly encouraging.   
}

\titlerunning{Monochromatic amplitude and phase variations for EC 20338$-$1925}
\authorrunning{S.K. Randall et al.}
\maketitle

\section{Introduction}    

Understanding the formation of subdwarf B (sdB) stars is one of the remaining challenges in stellar evolution theory today. As compact, low-mass, evolved objects located on the Extreme Horizontal Branch (EHB) it is generally accepted that they descend from the red giant branch, but loose too much of their hydrogen-rich envelope to ascend the asymptotic giant branch after core He-exhaustion \citep[for a comprehensive recent review on subdwarf B stars see][]{heber2009}. However, the details surrounding the mass loss in particular are debatable, and a number of evolutionary scenarios have been modelled \citep[e.g.][]{dorman1993,han2002,han2003,yu2009}. These include both single star and binary formation channels such as common envelope evolution, stable and unstable Roche lobe overflow, and the merger of two Helium white dwarfs. Each of these channels should leave a distinct signature on the internal composition and the mass distribution of the resulting sdB star population. While the latter may be computed dynamically for the very rare case of an eclipsing binary, the former can be probed only on the basis of asteroseismology.
 
Over the last decade, the asteroseismological exploitation of subdwarf B stars has become increasingly successful. Full asteroseismological analyses leading to a precise estimate of the mass and hydrogen envelope thickness have now been presented for ten rapidly pulsating sdB stars \citep[see, e.g.][for the most recent results]{charp2008,val2008b,randall2009}. These objects, also known as EC 14026 or V361 Hya stars after the prototype \citep{kilkenny1997} exhibit low-order, low-degree $p$-mode oscillations on a typical time-scale of 100$-$200~s, thought to be excited by a classical $\kappa$-mechanism associated with a local overabundance of iron in the driving region \citep{charp1996,charp1997}. While the asteroseismological results are beginning to indicate first evolutionary trends \citep[see][for a review]{charp2009}, it is of course highly desirable to verify the accuracy of the claimed structural parameters using an independent means. 

One of the most promising ways of doing this is to (partially) identify the modes of pulsation observationally. In the currently employed ``forward method'' in asteroseismology, the observed periodicities are fit to those computed for a grid of models, and an ``optimal'' model is identified using $\chi^2 $ minimisation. During the period fitting process, the observational frequencies are naturally associated with a theoretical mode of a given degree index $\ell$ and radial order $k$ (unless there is evidence for relatively fast rotation in the target, the computed period spectra are degenerate in azimuthal index $m$). Therefore, any independent knowledge of the modal indices can be used to reject an ``optimal'' model or, more usefully, be employed as an a priori constraint for the exploration of model parameter space. In certain cases, such constraints may be necessary to isolate just one ``optimal'' model from a number of very different possible solutions \citep{randall2009}, in others they may simply confirm the original solution \citep{val2008b}.  

The degree index of a mode can be observationally inferred using several techniques. For the special case of a rapid rotator, the presence or absence of rotational splitting can give clues to the $\ell$ value of at least the higher amplitude modes \citep{charp2005b,charp2008b}. Other, more universally applicable methods rely on the wavelength-dependent behaviour of the target flux during a pulsation cycle. Several recent projects have focussed on the interpretation of line profile variations observed in EC 14026 stars from high-resolution time-series spectroscopy \citep{telting2008,maja2009,telting2010}, but these efforts were always hampered by the low S/N of the data. Indeed, the relative faintness of subdwarf B stars ($V\gtrsim$ 12) coupled with the short exposure times acceptable ($\lesssim$ 30~s) for these fast pulsators make this type of study a challenge even when granted access to the world's largest telescopes, and none of the observations obtained to date were of sufficient quality to yield a unique identification of $\ell$. Meanwhile, numerous studies based on low-resolution time-series spectroscopy of EC 14026 stars \citep[e.g.][to name just a few]{jeffery2000,otoole2005,telting2006} have focussed mainly on measuring the radial velocity and equivalent width variations, and deducing the apparent changes in effective temperature and surface gravity over a pulsation cycle. Any attempts at mode identification were unfortunately inconclusive. 

A more promising approach is to analyse the amplitude and phase variations of a pulsation mode as a function of wavelength. This has been done quite successfully on the basis of multi-colour photometry. While the S/N of the observations is still a major limitation, particularly for fainter targets and/or low-amplitude pulsations, unique identifications of the degree index $\ell$ have been possible in several cases \citep{jeffery2004,randall2005,charp2008b,baran2008}. By far the nicest results to date were obtained for the brightest EC 14026 star Balloon 090100001 on the basis of UBV light curves of extraordinary quality obtained with the three-channel photometer LaPoune at the CFHT. Not only was the dominant pulsation clearly found to be a radial mode, but a further 8 frequencies could additionally be unambiguously identified as $\ell$=1, $\ell$=2 or $\ell$=4  modes \citep{charp2008b}. Beyond its intrinsic interest, this result proved beyond doubt the validity and accuracy of the theoretical colour-amplitudes calculated according to the method outlined in \citet{randall2005}.

The analysis presented here makes use of the same theoretical framework and numerical tools as our previous studies \citep{randall2005,charp2008b}, but it is based on time-series spectrophotometry rather than multi-colour photometry. To our knowledge this is the first time that this method of mode identification has been applied to a pulsating subdwarf B star. While the idea had been present in the back of our minds since the development of the theoretical tools, it was the very promising results presented by \citet{vankerkwijk2000} and \citet{clemens2000} for the ZZ Ceti white dwarf G29$-$38 that finally motivated us to apply for the necessary observing time. In that study, just under five hours worth of Keck LRIS time-series spectrophotometry were sufficient to measure the degree indices for six modes. Of course, G29$-$38 is an exceptionally bright ($V\sim$ 13) white dwarf with several relatively high amplitude ($A\gtrsim$ 1\%) periodicities, and the monochromatic flux behaviour of $g$-mode pulsations in ZZ Ceti stars is somewhat different to that of $p$-modes in EC 14026 stars \citep[cf.][]{randall2005,brassard1995}. Also, the strong broadening of the Balmer lines in white dwarfs greatly facilitates the observational characterisation of the flux behaviour across the line profiles, as do the longer pulsation periods and correspondingly higher integration times acceptable. Thus it was with contained optimism that we embarked on the present study. 

Given that the observations envisaged would be obtained with the Very Large Telescope (VLT) run by ESO on Cerro Paranal, Chile, we selected a target located in the southern hemisphere. EC 20338$-$1925 is a moderately bright ($V\sim$ 15.67) sdB star with no photometric indication of a cool companion \citep{kilkenny2006}, and atmospheric parameters around $T_{\rm eff} \sim$ 35,500~K, $\log{g}\sim$ 5.75, and $\log{N(\rm He)/N(\rm H)}\sim - $ 1.71 \citep{ostensen2010}. It was discovered to belong to the class of rapid EC 14026 variables by \citet{kilkenny2006}, who detected five periodicities near 168, 151, 147, 141 and 135~s on the basis of white-light photometry gathered in 1998. The period spectrum was very clearly dominated by the 147-s oscillation, which exhibited a high amplitude of nearly 2.5\%. This for an EC 14026 star unusually strong oscillation mode was the main reason we chose EC 20338$-$1925 over other, brighter target candidates. After the observing time had already been granted we learned that subsequent observations obtained from 1999 to 2007 had revealed quite striking amplitude variations for some of the modes. In particular, the 147-s oscillation had intermittently been measured at an amplitude as low as 0.3\% \citep{Kilkenny2010}, while some other frequencies appeared to have constant strengths. Such amplitude variations have been observed in a number of EC 14026 stars, however their origin is unclear, making EC 20338$-$1925 an enigmatic target for detailed study.

\section{Observations and Data Reduction}

\begin{table}[t]
\caption{Time-series spectroscopy obtained for EC 20338$-$1925}
\label{obslog}
\centering
\begin{tabular}{c c c c c}
\hline
\hline
Night & Date (UT) & Start time (UT) & Length (h) &  spectra pairs \\
\hline 
1 & 2009-08-22 & 23:57 & 04:48 & 654 \\
3 & 2009-08-25 & 00:18 & 04:27 & 615 \\
4 & 2009-08-25 & 23:48 & 04:52 & 656 \\
\hline
\end{tabular}
\end{table}

The observations presented here were obtained using the HIT-MS mode of FORS2 mounted on Antu (UT1) at the VLT. This rarely used mode is ideal for fast time-series spectroscopy as it allows 42 observations per readout, achieved by exposing only a small part of the chip and successively shifting the exposed blocks of columns across the CCD while already integrating on the next spectrum. Moreover, up to two closely spaced targets can be observed, allowing for both a main target and a comparison star to be monitored simultaneously. For more detailed information on FORS2 and the HIT mode please see the User Manual, available online via the ESO webpages.

We were allocated four consecutive half-nights from 22-25 August 2009 for our proposed study of EC 20338$-$1925. Making use of the blue-efficient E2V CCD formerly installed on FORS1 together with the cross-disperser grism 600B we hoped to obtain spectra covering the 3400-6500~\AA\ wavelength range. Unfortunately, we ran into severe technical problems at the beginning of the second night of observations, as one of the E2V chips of the mosaic died. Since in the HIT-MS mode each spectrum is projected along a block of columns spread across the two chips, this resulted in the loss of half the spectral wavelength range for the remainder of the observing run. To make matters worse, the mask we had cut before the beginning of the run placed our target on the dead chip, meaning that we could not resume our observations until a new mask had been cut the next day. Thus, despite good atmospheric conditions throughout the run we lost the second night completely, and obtained only half the wavelength range expected on the third and fourth nights. For a log of the observations please see Table \ref{obslog}.      

Since the main aim of our study was to examine the monochromatic properties of the target flux as a function of time, we avoided slit losses by choosing the largest mask openings available in HIT-MS mode (5$\times$5\arcsec), and essentially obtained spectrophotometry as the seeing was always well below the slit width. As a natural side effect, the spectral resolution of our data is not fixed but seeing-limited, and varies as a function of time. The typical seeing on the three nights we obtained data was $\sim$1.1\arcsec at $\sim$4500~\AA, which with the 600B cross-disperser corresponds to a wavelength resolution of $\sim$6.5\AA. On the first night (when both chips were working) we covered the 3500-4950~\AA\ (chip 2) and 5100-6600~\AA\ (chip 1) ranges, and opted for the bluer 3700-5200~\AA\ regime after chip 2 failed. That way, we included the Balmer lines from H$\alpha$ upwards on night 1, and from H$\beta$ upwards on nights 3 and 4 for our target star. We also monitored a F7V comparison star, HD 196286, picked purely for its location (7.5\arcmin\ from the target) and brightness ($V\sim$ 10.1, from Simbad). The spectral range covered for the comparison star is shifted $\sim$400~\AA\ bluewards with respect to the target star due to its different location on the chip. We set the exposure time for each pair of spectra to 25~s, yielding a total integration time of 1025~s (for 41 measurements) on the CCD before each 36~s readout. This means that only 0.3\% of the time-series is taken up by dead time in between exposures! 

\begin{figure}[t]
\centering
\includegraphics[width=7.0cm,angle=0,bb=170 240 470 630]{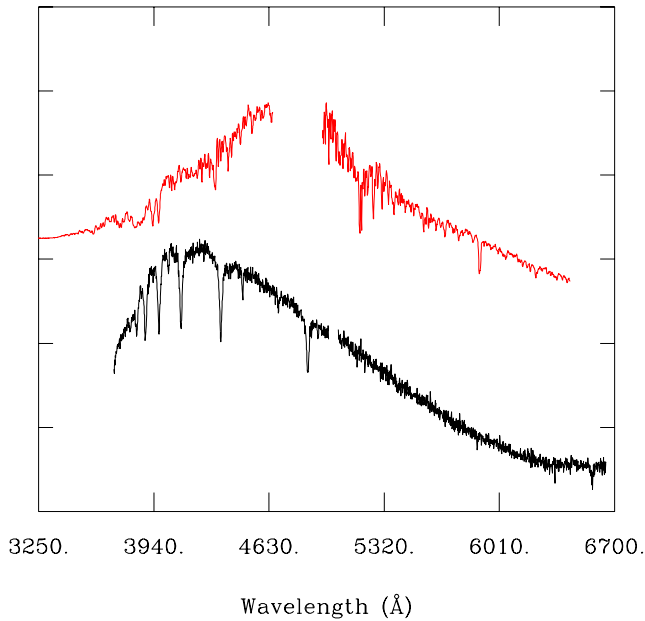}
\caption{Typical individual spectra obtained on 22 August 2009 for EC 20338$-$1915 (bottom) and the main sequence comparison star (top). The spectra have been arbitrarily normalised for visualisation purposes, and are not flux calibrated.}
\label{spectra}
\end{figure}

Given that the HIT-MS mode is not supported by the FORS pipeline, we wrote our own IRAF script to batch-process the data. This script essentially treats the 82 (41 each for the target and comparison star) spectra on each image as echelle orders, and writes the reduction products out as individual FITS files with a time-stamp corresponding to the middle of the exposure calculated for each pair of spectra. The steps performed are bias subtraction, bad pixel interpolation, cosmic ray cleaning, background fitting and subtraction, tracing and extraction of the spectra, and finally wavelength calibration. This last step was somewhat challenging, since we did not manage to take useful arc-lamp calibration data with the available wide-slit masks. In the end, the only way to obtain a meaningful wavelength scale was to use the spectral lines of the target and comparison star themselves, and compute the wavelength solutions for each night on the basis of the respective time-averaged combined spectra. Luckily, subdwarf B stars show a number of easily identifiable Balmer and Helium lines at short wavelengths, and we were thus able to obtain a reasonably accurate (within a few~\AA) wavelength calibration for at least the blue part of the target spectrum. This is sufficient for our purposes, since we do not attempt to detect radial velocity variations arising from the pulsations due to the low resolution of the spectra and the arbitrary wavelength variations induced by spatial shifts of the target across the wide slit. The red part of the target spectrum is largely devoid of features, and the wavelength solution obtained cannot be trusted for quantitative analysis. For the comparison star, we were able to identify several lines on both chips, and achieve a rough wavelength calibration. However, since these measurements are only used in wavelength-integrated form in what follows, this calibration is needed as nothing more than a guideline.

Figure \ref{spectra} shows a typical individual spectrum for both the target and comparison star, taken from the first night where both chips were available. Note that the comparison star data were chopped at the short-wavelength end of the red chip as this part of the spectrum was saturated. The high quality is striking considering the relative faintness of the target \citep[V$\sim$15.67 from][]{kilkenny2006} and the short exposure time used (25~s). In the central part of the blue target spectrum, we measure a signal-to-noise S/N $\sim$ 55, while it is lower in the red part at S/N $\sim$ 25. For the comparison star an accurate measurement of the S/N is complicated by the numerous metal lines, but we can roughly estimate S/N $\sim$ 90 for the central part of the blue spectrum and S/N $\geq$ 100 for the red part.  It is apparent from looking at the relative depth of $H\alpha$ compared to the other Balmer lines in the target spectrum that the red part is a lot less useful than the blue part in terms of S/N of the lines, mostly due to the intrinsically low flux of sdB stars at longer wavelengths. Adding to this the fact that we were able to obtain only half a night's worth of red data, and that an accurate wavelength calibration was not possible, they become unsuitable for a detailed interpretation. Therefore, the analyses presented in the following sections are based on the blue data only.   

\section{Data Analysis}

\begin{figure}[b]
\centering
\includegraphics[width=7.0cm,angle=0,bb=120 240 500 630]{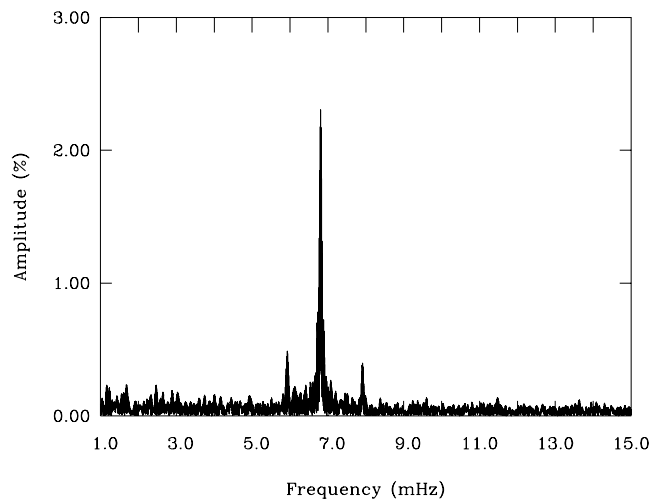}
\caption{Fourier transform computed on the basis of the integrated flux from all blue spectra of sufficient quality gathered during the observing run.}
\label{ft}
\end{figure}

\subsection{Frequency Determination}

\begin{table}[t]
\caption{Periodicities extracted for EC 20338$-$1925. We list the period and amplitude obtained from our data set, as well as the ranges of amplitude measured by D. Kilkenny from 1998$-$2007. The uncertainty on our amplitudes is $\sim$0.05\% from the least squares fit to the light curve.}
\label{freqs}
\centering
\begin{tabular}{c c c c}
\hline
\hline
Rank & Period (s) & Amp (\%) & A$_{Kilkenny}$(\%) \\
\hline 
1 & 146.940 & 2.30 & 0.32$-$2.40 \\
2 & 168.435 & 0.45 & 0.33$-$0.46 \\ 
3 & 126.382 & 0.38 & 0$-$0.33 \\
4 & 140.588 & 0.20 & 0$-$0.24 \\
\hline
\end{tabular}
\end{table}

\begin{figure}[b]
\centering
\includegraphics[width=6.5cm,angle=0,bb=120 170 500 680]{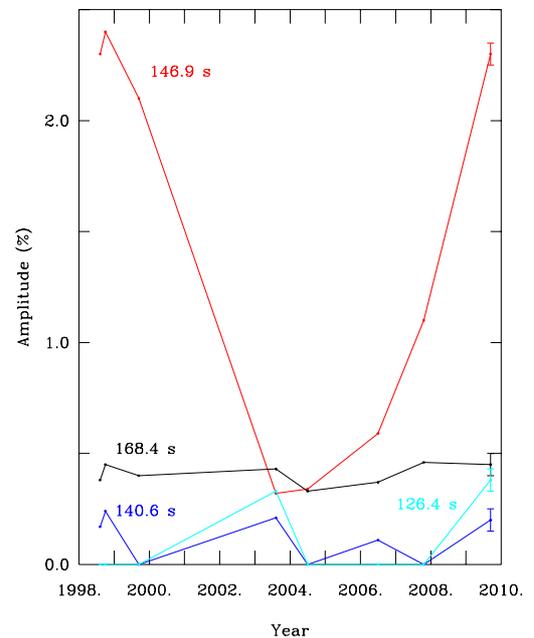}
\caption{Measured amplitudes of the 4 periodicities listed in Table \ref{freqs} over the last 12 years. Apart from the most recent data points, which were taken from the observations presented here, the measurements were kindly made available to us by D. Kilkenny. The errors were not provided, but should be similar to or smaller than those shown for our data. Zero amplitudes imply that the pulsation was not detected in a particular data set down to a threshold of $\sim$0.1\%.}
\label{amps}
\end{figure}

The first step in the data analysis was to obtain pulsation frequencies for our data set, and at the same time assure the quality of the measurements retained for further analysis. In order to maximise the S/N we integrated the flux of each spectrum over the wavelength range of interest (3650$-$4950~\AA) and produced broadband fluxes for the target as well as the comparison star. These were then used to compute normalised differential light curves additionally corrected for differential extinction by fitting a second-order polynomial to each night's data. Individual outliers (usually associated with bad seeing) were removed from the light curves prior to Fourier analysis, and the corresponding spectra were discarded. While we initially included the red data, we found that this gave results of lower quality than using the blue part of the spectra alone, and thus decided to consider only the latter. In addition, the last $\sim$hour of data taken on the first night had to be disregarded because the stars had shifted towards the edge of the slit and some flux loss had occurred, making these measurements unsuitable for time-series flux analysis. 

We show the Fourier transform (FT) of the combined light curve in Fig. \ref{ft}. By pre-whitening we were able to detect four periodicities above the imposed 4$\sigma$ threshold for credible pulsations (0.2\% of the mean brightness), listed in Table \ref{freqs} together with their amplitudes. Three of these, including the  strong dominant pulsation (146.9~s), correspond well to those measured in 1998 \citep{kilkenny2006}, while the fourth (126.4~s) was previously only detected from a dataset taken in 2003 \citep{Kilkenny2010}. As briefly mentioned in the Introduction, the amplitudes of some of the periodicities observed for EC 20338$-$1925 (particularly the dominant 146.9-s peak) are highly variable on a time-scale of several years, whereas others (such as the 168.4-s pulsation) show a remarkably stable amplitude over the course of the last decade. This is visualised in Fig. \ref{amps}, where we have plotted the amplitudes measured by D. Kilkenny during 7 observing runs between 1998 and 2007 for the four oscillations detected from our data together with our more recent results (see also the last column of Table \ref{freqs}). It is not clear whether these observed amplitude variations are due to real intrinsic amplitude modulations, or caused by the beating of very close unresolved modes. We hope to shed some light on this in the following sections. 

\subsection{Atmospheric analysis}

\begin{figure}[t]
\centering
\includegraphics[width=6.5cm,angle=270,bb=100 150 520 650]{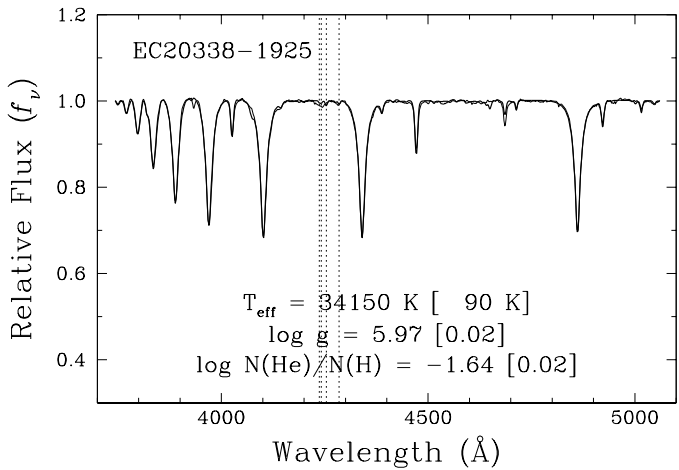}
\caption{Normalised averaged spectrum for EC 20338$-$1925 (thin line) overplotted with the best synthetic spectrum ``with metals'' (thick line). The vertical dashed lines indicate the positions of the SIII, OII and NII lines thought to be present in the observed spectrum. The atmospheric parameters derived from this fit are also given.}
\label{metalspec}
\end{figure}

\begin{figure*}[t]
\centering
\begin{tabular}{cc}
{\includegraphics[width=6.8cm,angle=270.0,bb=60 80 570 730]{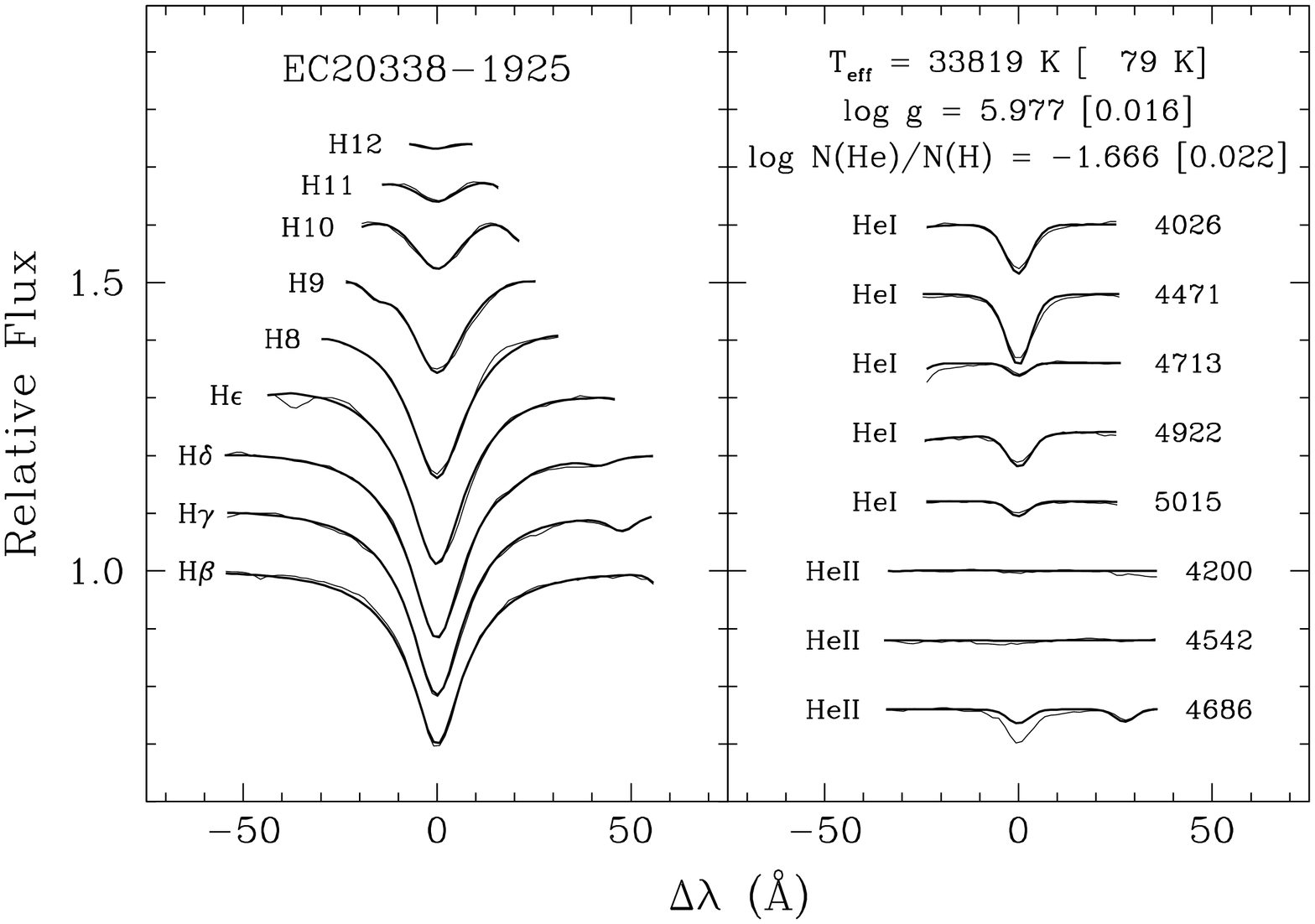}} & {\includegraphics[width=6.8cm,angle=270.0,bb=60 80 570 730]{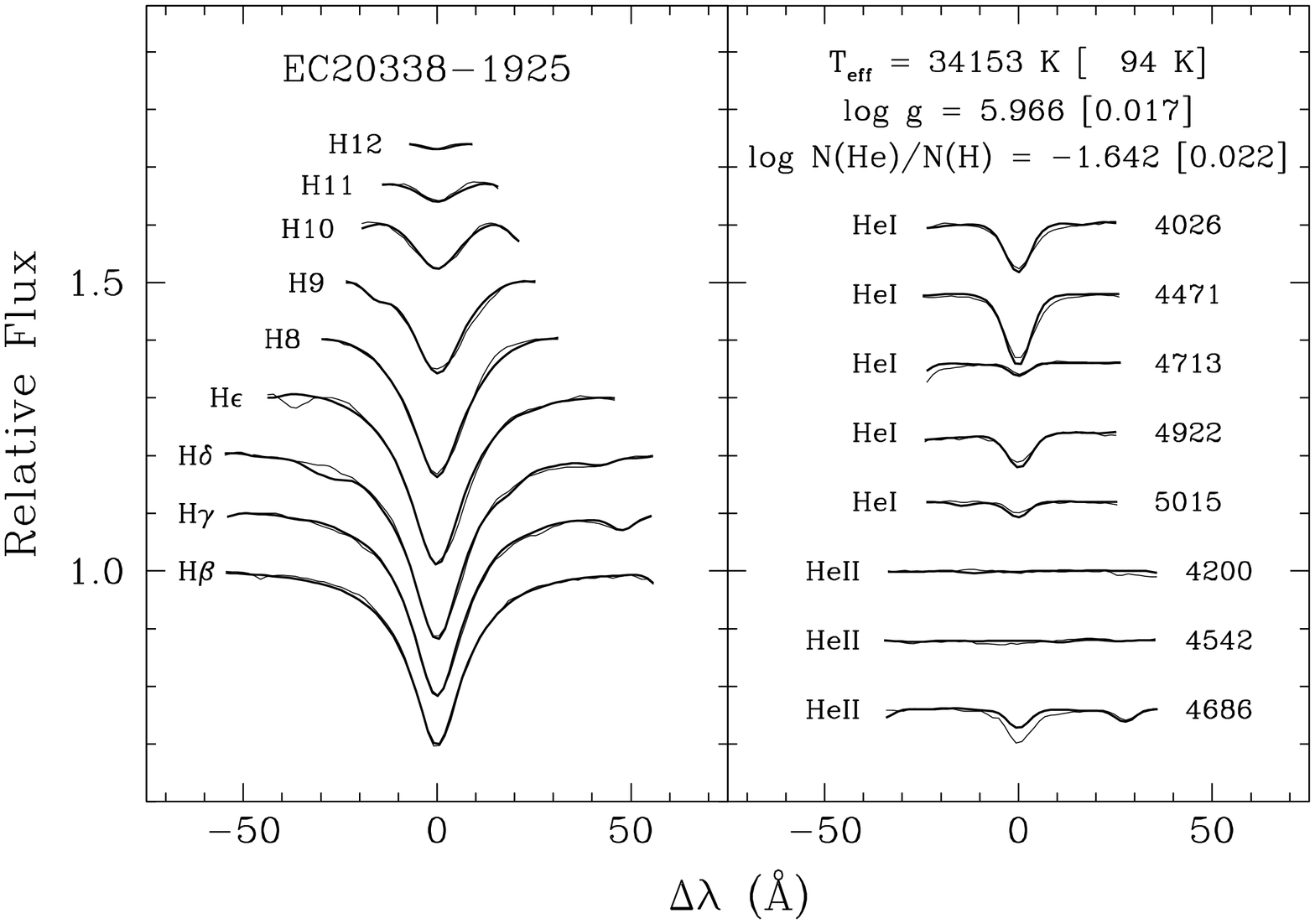}} \\
\end{tabular}
\caption{{\it Left panel:} Available H and He lines from the combined FORS2 spectrum of EC 20338$-$1925 (thin lines), overplotted with the best atmosphere fit not including metals (thick lines). The atmospheric parameters derived are indicated in the plot. There are signs of interstellar absorption as revealed by the Ca II K feature in the blue wing of H$\epsilon$. {\it Right panel:} The same as the left panel, except that here metals are included in the proportions listed in the text.}
\label{fitboth}
\end{figure*} 

The second step in the data analysis was to determine the atmospheric parameters of EC 20338$-$1925, which are necessary for the modelling of the monochromatic flux behaviour (see below). For this purpose, we made a combined spectrum on the basis of 995 (blue) spectra spread over the three nights that have central wavelength FWHMs close to the average seeing during the observing run (between 1.0$-$1.15$\arcsec$). The result was a time-averaged spectrum with wavelength resolution 6.4$\pm$0.2~\AA\ and high S/N of $\sim$400 at the central wavelength.

We fit this averaged spectrum with two grids of synthetic spectra computed from NLTE model atmospheres specially designed for subdwarf B stars using the public codes TLUSTY and SYNSPEC \citep{hubeny1995,lanz1995}. The spectra in the first grid contain only H and He in the atmospheric composition (``no metals''), while those of the second set (``with metals'') have a chemical composition made up of H, He, C (0.1 solar), N (solar), O (0.1 solar), S (solar), Si (0.1 solar) and Fe (solar). This mixture was derived from the work of \citet{blanchette2008}, who determined the abundances of several astrophysically important elements for five representative subdwarf B stars on the basis of FUSE spectra in conjunction with NLTE model atmospheres. The abundance patterns were found to be very similar in all cases, therefore we deemed it appropriate to include the most abundant heavy elements in our model atmospheres at the level measured.    

Fig. \ref{fitboth} shows the fits to the Balmer and Helium lines obtained for both the ``no metals'' and ``with metals'' cases, and also gives the inferred values for the effective temperature, logarithmic surface gravity and logarithmic fractional Helium abundance. The uncertainty estimates given of course refer only to the formal fitting errors and most certainly underestimate the real errors. As is typically the case for EC 14026 stars with $T_{\rm eff}\sim$35,000~K, the resulting atmospheric parameters are only marginally influenced by the inclusion of metals in the model atmospheres, and are consistent with each other \citep{heber2000,geier2007}. And indeed, the line profile fits shown in Fig. \ref{fitboth} hardly change from one plot to the next. 

There are nevertheless some indications that the solution incorporating metals is to be preferred. Despite its relatively low resolution, the averaged spectrum of EC 20338$-$1925 does hint at the presence of a few heavy elements, thanks to the high quality of the data. In particular, the normalised spectrum plotted in its entirety in Fig. \ref{metalspec} contains lines corresponding to absorption from NII (4237.05 and 4241.79~\AA), OII (4253.87~\AA) and SIII (4284.97~\AA). These are marked by vertical dotted lines in the plot, and are recovered quite nicely by the overplotted ``with metals'' synthetic spectrum. Another indicator for the presence of metals in EC 20338$-$1925 is the relatively poor fit to the He II line at 4686~\AA\ (see Fig. \ref{fitboth}). While the model incorporating metals fares better in the matching of this line than the synthetic spectrum without metals, it appears that even the ``with metals'' chemical composition does not include enough metals to accurately represent EC 20338$-$1925. Given the fact that the spectroscopic solution is not very sensitive to the metallicity of the model atmosphere for our target, this does not constitute a problem for the accuracy of the inferred atmospheric parameters. Therefore, we adopt the ``with metals'' values of $T_{\rm eff}$=34,153~K and $\log{g}$=5.966 in what follows. 

\subsection{Observed monochromatic amplitudes and phases}

\begin{figure}[t]
\centering
\includegraphics[width=7.5cm,angle=0,bb=120 165 480 670]{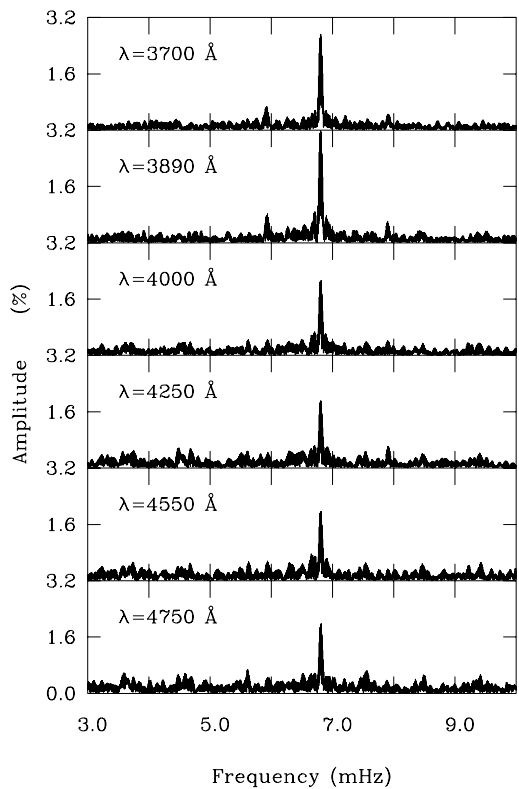}
\caption{Fourier transforms obtained for 10~\AA\ wavelength bins at the positions indicated in the left hand panel of Fig. \ref{ampphase}.}
\label{multifts}
\end{figure}

\begin{figure*}[t]
\centering
\begin{tabular}{cc}
{\includegraphics[width=7.5cm,angle=0,bb=100 200 550 670]{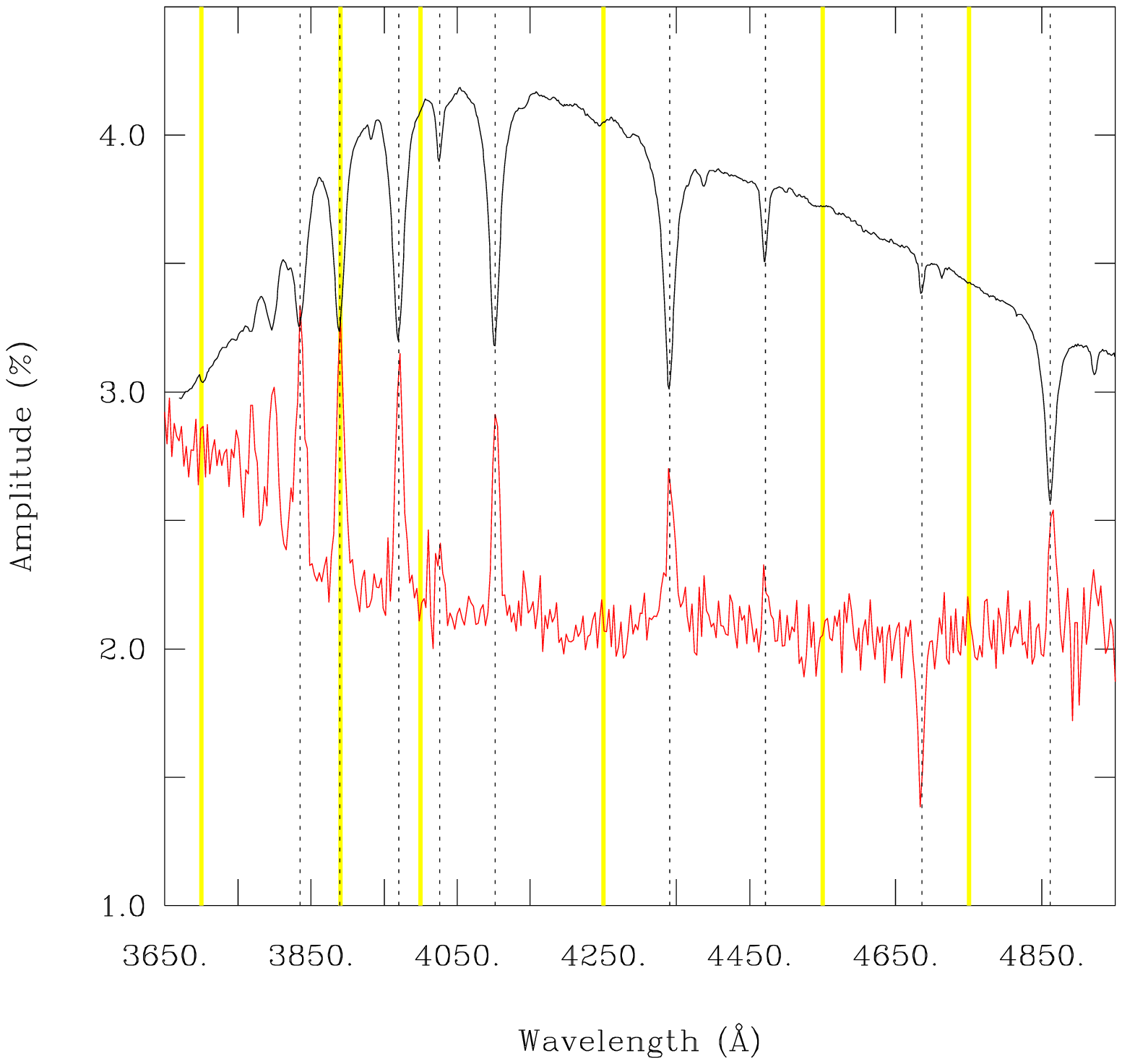}} & {\includegraphics[width=7.5cm,angle=0,bb=100 200 550 670]{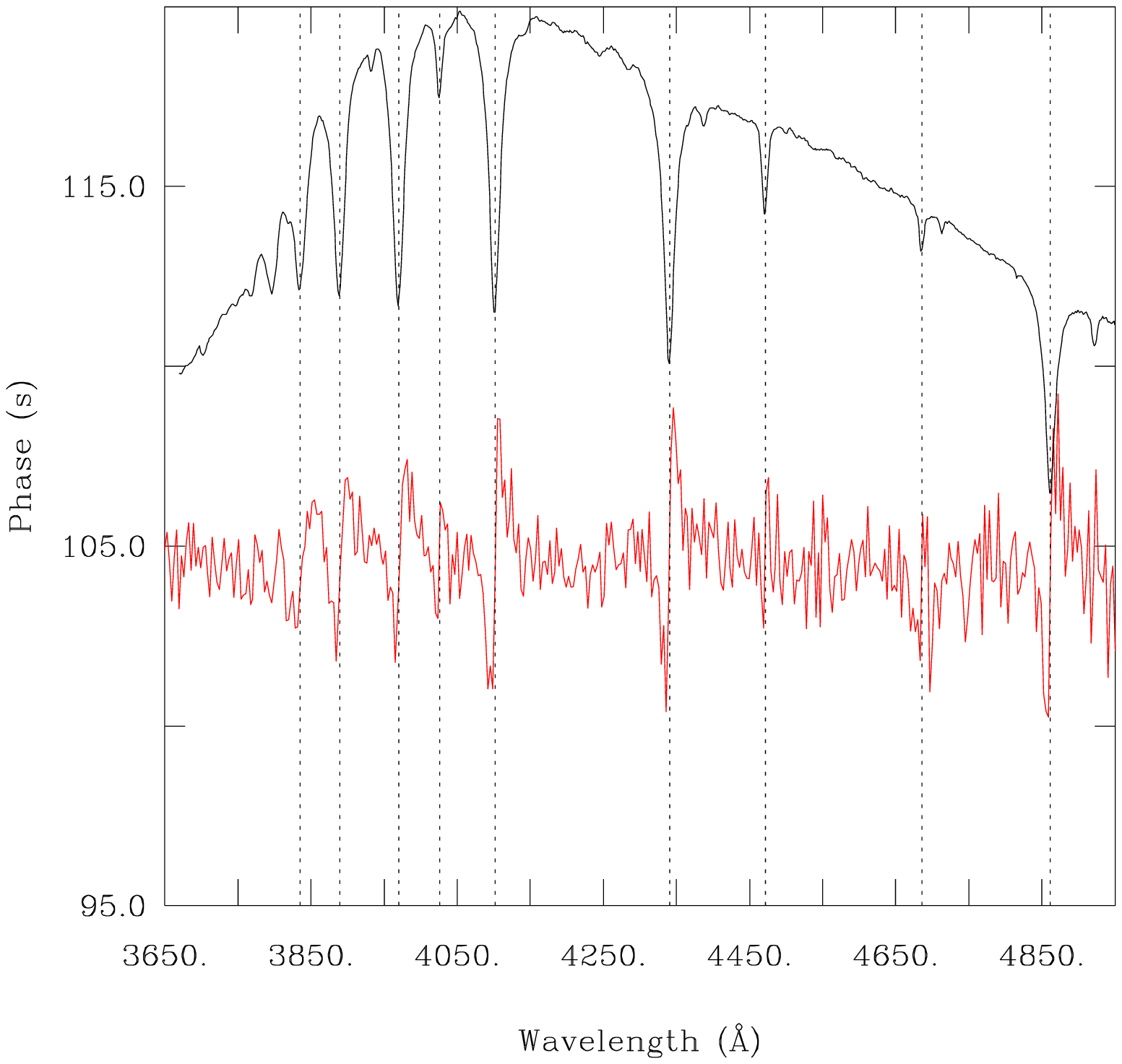}} \\
\end{tabular}
\caption{{\it Left panel:} We show the observed amplitude as a function of wavelength for the dominant pulsation mode $f_1$. For reference we plot the averaged spectrum of EC 20338$-$1925 on an arbitrary y-axis scale and also indicate the central wavelengths of the Balmer and prominent He lines (dotted vertical lines). The thicker yellow bars mark the wavelength bins for which the Fourier Transforms plotted in Fig. \ref{multifts} were computed. {\it Right panel:} The same as the left panel, but for the observed phase.}
\label{ampphase}
\end{figure*}

The final, and arguably most interesting part of the data analysis was the computation of the observed monochromatic amplitudes and phases for the periodicities extracted. For this exercise we used only the blue target spectra also employed for the computation of the broadband Fourier transform shown in Fig. \ref{ft}, thereby ensuring a high data quality and the exact same sampling. The latter is important since we used the periods determined in Section 3.1 as input for the least-squares fitting procedure detailed below. All selected target spectra were chopped to the 3650$-$4950~\AA\ range of interest and binned into common 3.25~\AA\ wavelength units, which were chosen to adequately sample the average $\sim$6.5~\AA\ resolution of the original data. 

Light curves were then obtained for each of the 401 wavelength bins, and corrected for differential extinction using the comparison star data. In order to achieve the best possible S/N, we integrated the comparison star spectra from 3650~\AA\ to their upper limit of $\sim$ 4700~\AA\, and employed the resulting broadband fluxes for the calculation of the differential light curves. For each wavelength bin, and using the periods listed in Table \ref{freqs} as fixed input parameters, the corresponding amplitudes and phases were determined from a least-squares fit to the light curve assuming a sinusoidal signal. While this did not involve the computation of Fourier spectra, we nevertheless calculated them for a small number of selected wavelength points as a sanity check. For illustration purposes we show some of these Fourier transforms in Fig. \ref{multifts}. Note that here, the spectra were binned into units of 10~\AA\ to somewhat decrease the noise level.      

Fig. \ref{ampphase} shows the monochromatic amplitudes and phases derived for the dominant mode $f_1$. For reference we also show the averaged spectrum of EC 20338$-$1925, and indicate the rest wavelengths of the Balmer and most prominent Helium lines. Looking at the amplitude spectrum we find that the amplitude of pulsation in the continuum drops quite rapidly between 3650 and 4050~\AA\, and then stabilises somewhat. This was to be expected, since the same effect is observed from multi-colour photometry \citep[e.g.][]{charp2008,randall2005,jeffery2004}. In the Balmer line cores, the pulsational amplitude is significantly higher than in the continuum, and again decreases with increasing wavelength of the line. Considering the Helium lines strong enough to leave a visible imprint on the amplitude spectrum, it appears that the neutral helium lines at 4026 and 4471~\AA\ are associated with higher amplitudes compared to the continuum, while the pulsational signature is extremely weak in the He II line at 4686~\AA. The observed phase on the other hand appears to be constant in the continuum, jumps occuring only in the line centers. This is consistent with the very small phase shifts observed on the basis of multi-colour photometry \citep{jeffery2004,jeffery2005}.

\section{Modelling the observed monochromatic amplitude and phase variations}

\begin{figure}[t]
\centering
\includegraphics[width=6.5cm,angle=0,bb=120 165 480 670]{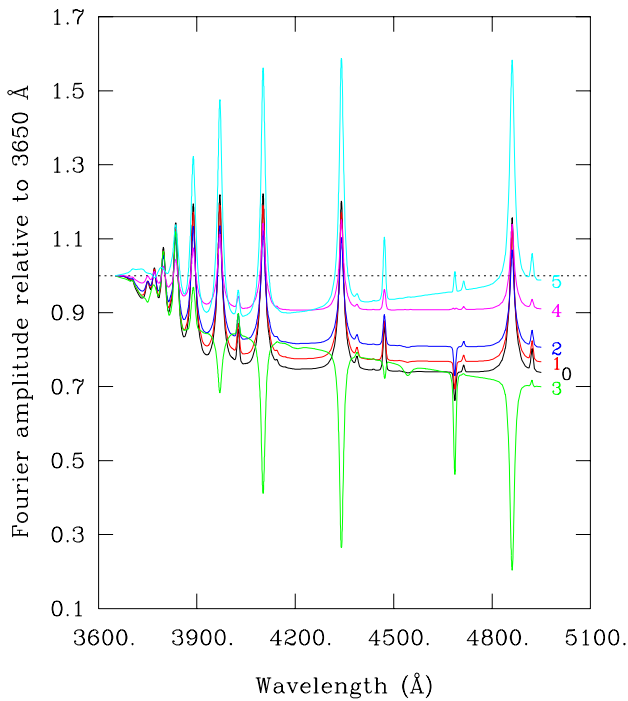}
\caption{Theoretical monochromatic amplitudes for modes with degree indices $\ell$=0-5 in the wavelength range of interest, normalised to 3650~\AA. The computations were specifically carried out for the dominant mode in EC 20338$-$1925, and assume $T_{\rm eff}=$ 34,153~K, $\log{g}=$ 5.966, $M_{\ast}/M_{\odot}=$ 0.48, $\log{q(\rm H)}=-$4.3, and $P=$ 146.9~s. The wavelength resolution was artificially degraded to 6.5\AA.} 
\label{theoryamps}
\end{figure} 

\subsection{Computation of theoretical amplitudes and phases}

For the computation of the theoretical monochromatic amplitudes and phases we closely follow the approach developed by \citet{randall2005} for the application to multi-colour photometry. We refer the interested reader to that paper for details on the underlying theory and the models, and only give a very brief overview of the methodology followed here. 

Our theoretical amplitudes and phases rely primarily on monochromatic quantities derived from model atmospheres. The computation of these quantities involves not only the standard specific intensities, but also the corresponding derivatives with respect to effective temperature and surface gravity across the visible disk of the star. They are calculated on the basis of a grid of LTE model atmospheres characterised by a uniform composition specified by $\log{[N(\rm He)/N(\rm H)]}=-$2.0, i.e. without metals.  Whereas it would be interesting to assess the effect that incorporating metals would have on this type of calculation in the future, the current grid is quite sufficient for the purposes of this study. In any case, the heavy element abundance pattern of EC 20338$-$1925 has not yet been characterised, and most likely deviates from the \citet{blanchette2008} composition employed for the atmospheric analysis detailed in Section 3.2. We therefore compute the monochromatic atmospheric quantities using the existing H/He model grid, assuming an effective temperature $T_{\rm eff}$=34,153~K and a logarithmic surface gravity $\log{g}$=5.966 as derived above.

While the monochromatic atmospheric quantities thus computed are the main contributors to the relative first order perturbations of the emergent flux (from which the monochromatic amplitudes and phases directly follow), they neglect to take into account non-adiabatic effects. The quantities describing the departure from adiabacity in amplitude and phase must instead be derived from full non-adiabatic pulsation calculations on the basis of stellar models. For this purpose, we employ our well-known ``second-generation'' models \citep[see, e.g.][]{charp1996,charp2001} in conjunction with adiabatic and non-adiabatic pulsation codes based on finite-element techniques \citep{brassard1992,fontaine1994}. These are exactly the same tools that have been used very successfully in the past for the asteroseismological analysis of EC 14026 stars. They require four structural input parameters: $T_{\rm eff}$, $\log{g}$, the total stellar mass $M_{\ast}$, and the fractional thickness of the hydrogen-rich envelope $\log{q(\rm H)}\equiv\log{M(\rm H)/M_{\ast}}$. In our computations we set the first two quantities to those inferred in Section 3.2, while for the other two we selected representative values of $M_{\ast}=$ 0.48 $M_{\odot}$ and $\log{q(\rm H)}=-$4.3. The two latter parameters were chosen to be typical of rapid sdB pulsators as found from asteroseismological studies, but do not significantly affect the results as long as they take on physically reasonable values.  

The quantities describing the departure from adiabacity were first computed for the period spectrum of the representative EC 20338$-$1925 model, and subsequently determined for the dominant 146.9-s dominant mode using interpolation techniques. While for $p$-mode pulsators these quantities are quite sensitive to the period of the oscillation in question, they are luckily not dependent on the degree index of the mode, and can thus be inferred quite easily for any period in the excited frequency range predicted by the model. For reference, at the period of the main mode the non-adiabatic coefficients are $R=$ 0.810 and $\Psi_T=$ 2.761 rad compared to their adiabatic values of $R$ = 1 and $\Psi_T=\pi$ \citep[for a definition of the parameters please see][]{randall2005}. With both the monochromatic atmospheric quantities and the non-adiabaticity coefficients available, we can now easily compute the monochromatic amplitudes and phases for EC 20338$-$1925.

\begin{figure}[t]
\centering
\includegraphics[width=6.5cm,angle=0,bb=100 70 480 730]{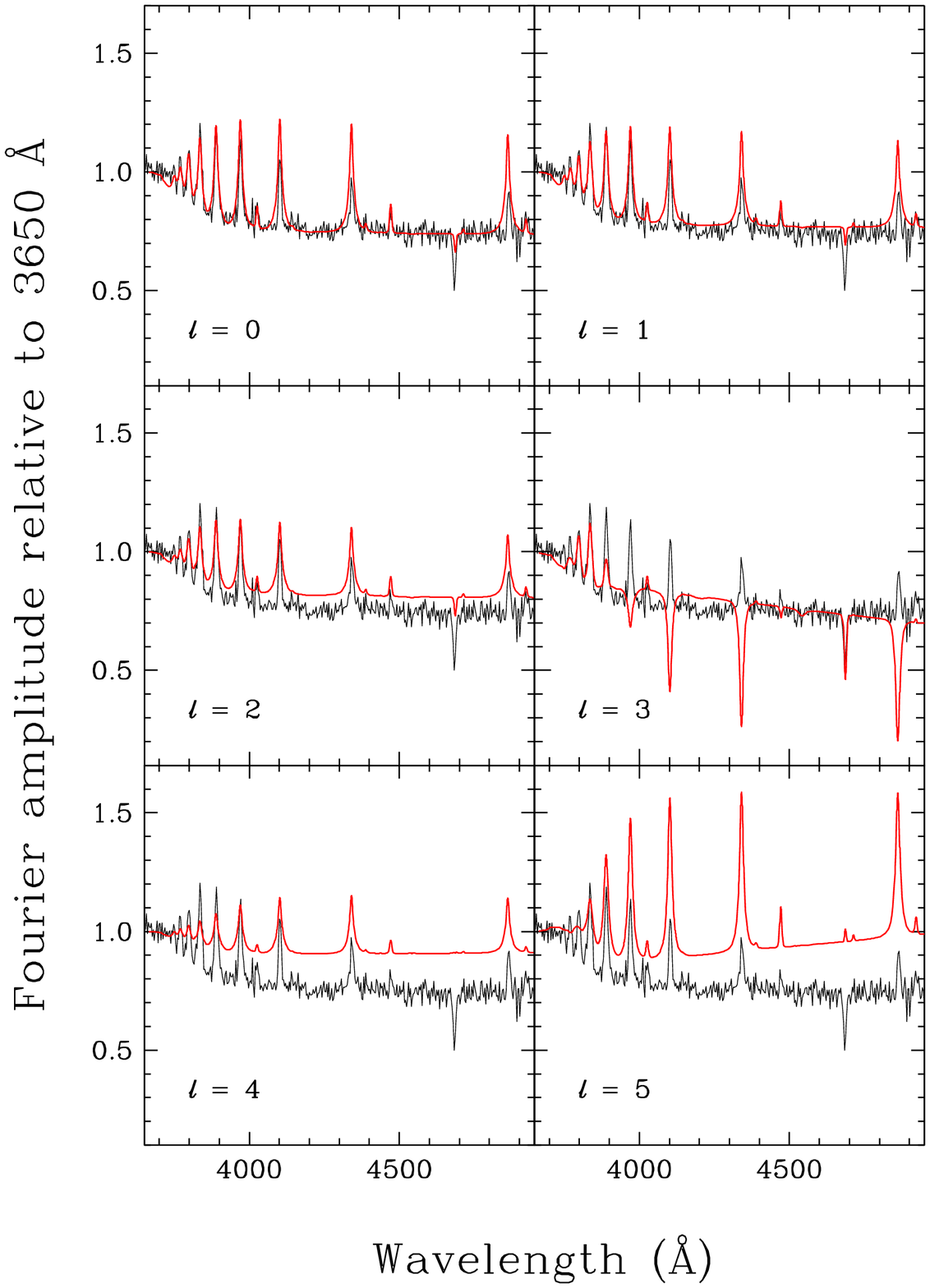}
\caption{Theoretical monochromatic amplitudes for modes with $\ell=$ 0-5 overplotted on the observed amplitudes for the dominant $f_1$ mode in EC 20338$-$1925. The curves have all been normalised to unity at 3650~\AA.} 
\label{ampsf1}
\end{figure}

\begin{figure*}[t]
\centering
\begin{tabular}{cc}
{\includegraphics[width=7.5cm,angle=0,bb=100 160 570 670]{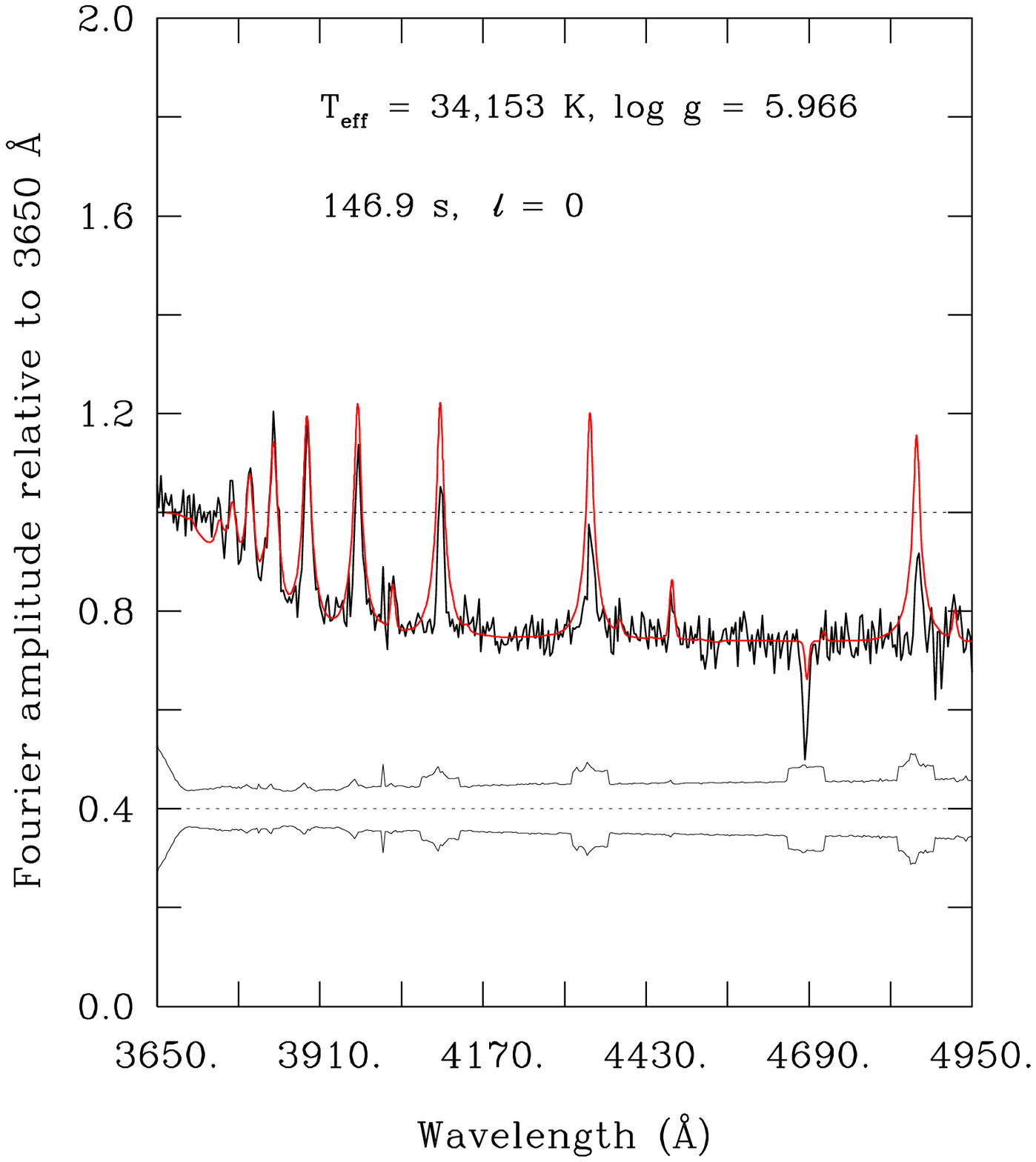}} & {\includegraphics[width=7.5cm,angle=0,bb=100 160 570 670]{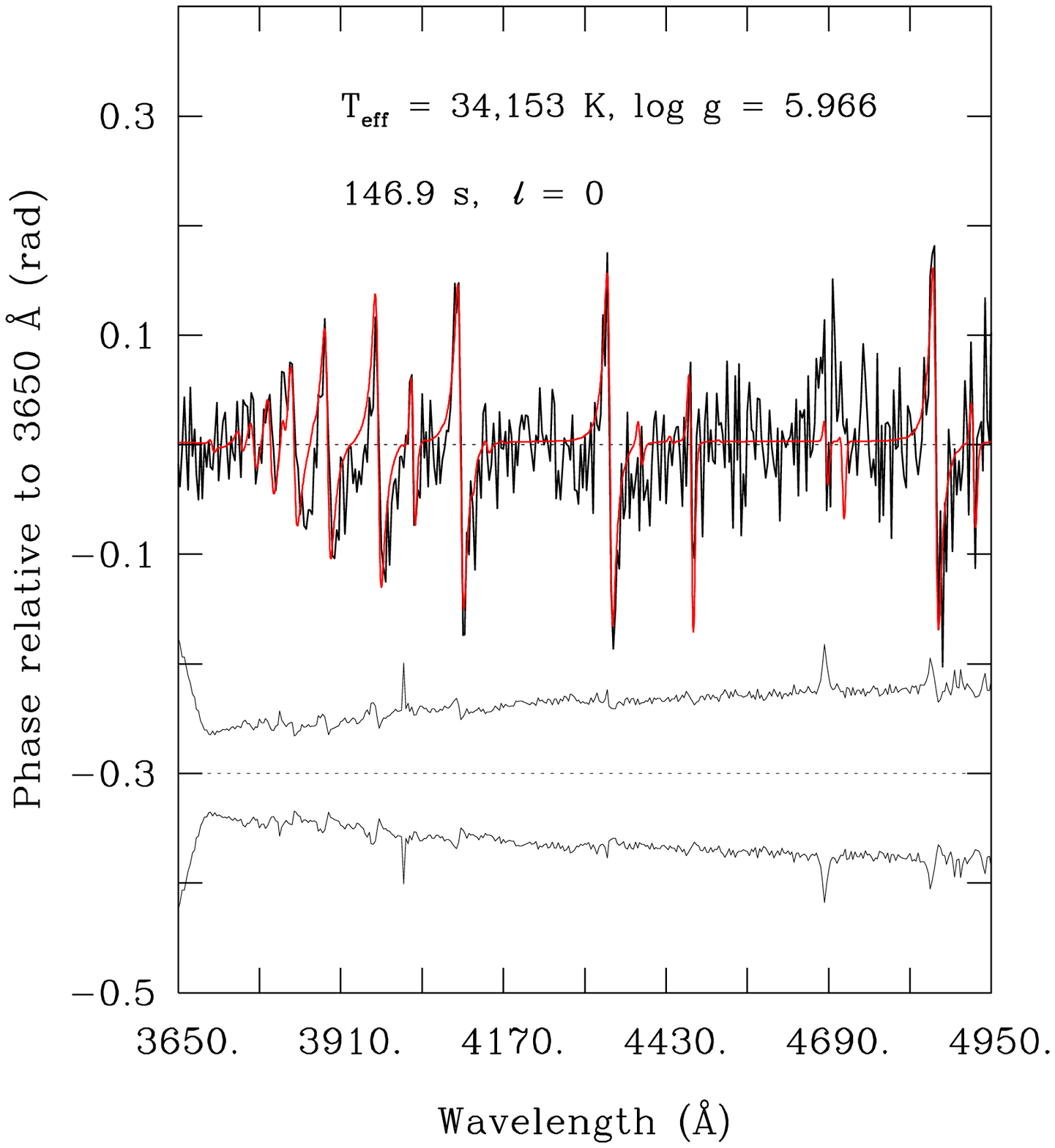}} \\
\end{tabular}
\caption{{\it Left panel:} We show the normalised observed amplitude for $f_1$ (black curve) together with the errors on the data points (thin curves shifted downwards along the y axis for clarity). The theoretical monochromatic amplitude as calculated for a radial mode with $P$=146.9~s and model parameters characteristic of EC 20338$-$1925 is overplotted (red curve). {\it Right panel:} Similar to the left panel, but referring to the monochromatic phase. The theoretical monochromatic phase was computed using the same parameters as for the amplitude, and additionally a radial velocity shift of 11~km/s was assumed.}
\label{zoomampphase}
\end{figure*}

Fig. \ref{theoryamps} shows the relative amplitudes as expected for the 146.9-s mode of EC 20338$-$1925 for degree indices from $\ell$=0 to $\ell$=5 in the wavelength range of interest. The curves have been normalised to unity at 3650~\AA\, and were degraded in wavelength resolution to the 6.5~\AA\ typical of the observational data. Higher order modes are not illustrated as they are thought to have extremely low amplitudes  in the continuum due to cancellation effects when integrating across the visible disk of the star, and should thus not be observable. As explained in \citet{randall2005}, the theoretical amplitudes and phases are, to first order, independent of the azimuthal index $m$, and can therefore be used for resolved or unresolved frequency components of multiplets split by slow rotation. The curves quite clearly show that the discrimination between low-order modes with $\ell\leq$ 2, and particularly between radial and dipole modes is challenging due to their very similar behaviour as a function of wavelength. This has always been the main difficulty also with the interpretation of multi-colour photometry and line-profile variability, and has limited many studies to the ambiguous identification of the main pulsations as low-order (i.e. $\ell$ = 0, 1 or 2) modes due to the insufficient S/N of the data \citep[e.g.][]{jeffery2005,tremblay2006,maja2009}. Given that the amplitude dependence is greatest at short wavelengths, it is therefore imperative to observe as far in the blue as possible. 

Comparing the curves displayed here for EC 20338$-$1925 to equivalent data for other EC 14026 stars \citep[see in particular Fig. 2 of][]{randall2005}, it turns out that we were quite lucky with our target in terms of discriminative power between the amplitudes of low-order modes. Indeed, as can be seen from Fig. 28a of \citet{randall2005}, the amplitude-wavelength behaviour of radial, dipole, and quadrupole modes diverges more strongly for stars with a higher surface gravity. By contrast, the $\ell$=3 mode shows a shallower amplitude gradient as a function of wavelength. EC 20338$-$1925 being one of the EC 14026 stars with the highest surface gravity measured is then definitely to our advantage, and increases the chances for the unique mode identification attempted below.

\subsection{Fitting the observed monochromatic amplitudes and phases}

We now compare the theoretical monochromatic amplitudes computed in the previous section to the observations. Fig. \ref{ampsf1} shows the curves displayed in Fig. \ref{theoryamps} overplotted on the observed monochromatic amplitude of the dominant $f_1$ mode, which has also been normalised to unity at 3650~\AA. From visual inspection alone it is obvious that the $\ell=$ 0 solution presents by far the best match to the observational data, followed by the curves for $\ell=$ 1 and, with a significant degradation $\ell=$ 2. The amplitudes for the higher degree modes match the data very poorly, and can be excluded from the outset. Computing the quality-of-fit $Q$ on the basis of $\chi^2$ fitting confirms this result, with $Q$=0.755 for $\ell=$ 0, $Q=$ 0.340 for $\ell=$ 1 and $Q=$ 0.056 for $\ell=$ 2. For the higher degree modes, $Q\ll$ 0.001, which implies that they can formally be excluded as possible solutions \citep{press1986}. The $Q$ value determined for the radial mode implies a very good fit, more than twice as good as that for the dipole mode. While the uncertainties on the observed amplitudes are too large to formally exclude the solutions for $\ell=$ 1 and $\ell=$ 2 as possible matches to the data, we believe that the far superior quality of the fit for $\ell=$ 0 leaves no doubt that the dominant pulsation observed in EC 20338$-$1925 is a radial mode. 

\begin{figure*}[t]
\centering
\begin{tabular}{cc}
\centering
{\includegraphics[width=8.5cm,angle=0,bb=20 70 580 740]{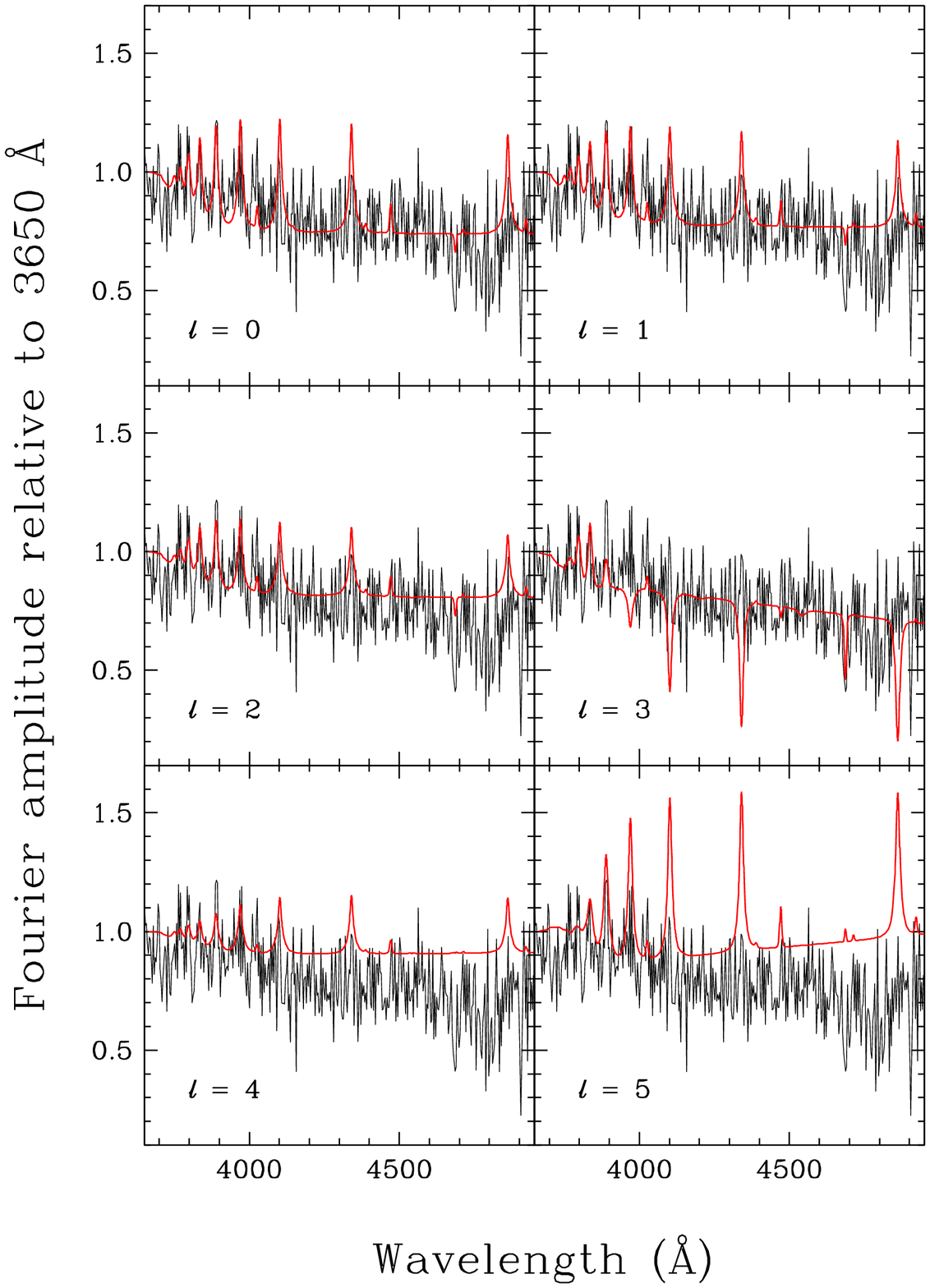}} & {\includegraphics[width=8.5cm,angle=0,bb=20 70 580 740]{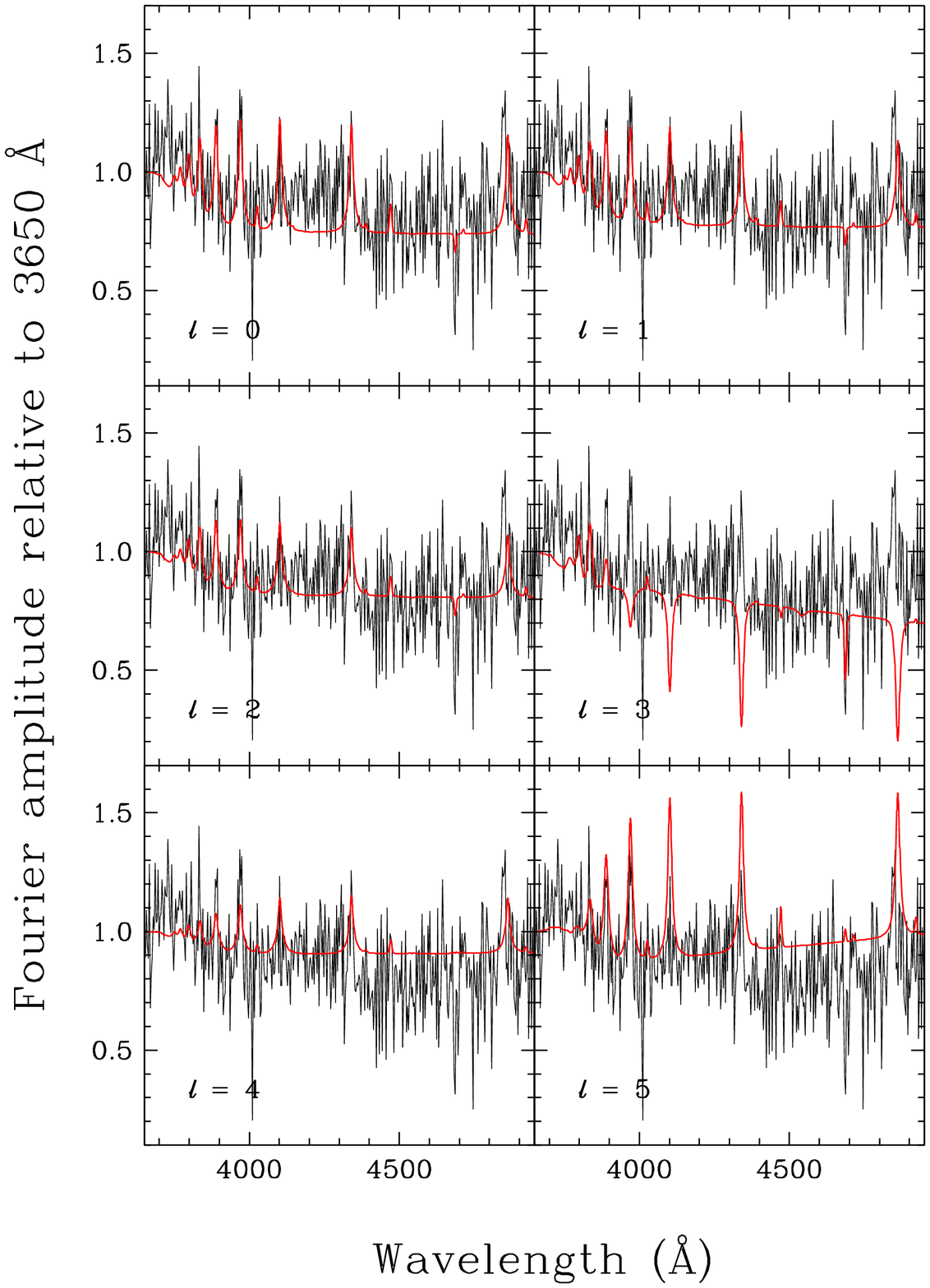}} \\
\end{tabular}
\caption{{\it Left panel:} Theoretical monochromatic amplitudes for modes with $\ell=$ 0-5 overplotted on the observed amplitudes for the $f_2$ periodicity in EC 20338$-$1925. The curves have all been normalised to unity at 3650~\AA. {\it Right panel:} Similar to the left panel, but referring to $f_3$.}
\label{ampsf2f3}
\end{figure*}

In order to illustrate the quality of the fit more clearly, we zoom in on the monochromatic amplitude plot for $\ell=$ 0 in the left panel of Fig. \ref{zoomampphase}. Here, we also include the uncertainty on the observed amplitudes as obtained from the least-squares fits to the light curve for each wavelength bin (thin vertical line segments shifted downwards along the y-axis for clarity). It can be seen that, while the observed amplitude behaviour in the continuum is matched nearly perfectly by the theoretical curve, there are some discrepancies in the spectral lines. Most strikingly, the amplitude dip corresponding to the He II line at 4686~\AA\ is predicted far weaker than observed, much like what we found for the model atmosphere fit to the time-averaged spectrum in Fig. \ref{fitboth}. This is almost certainly due to the lack of metals in the model atmospheres used for the computation. A more puzzling feature is that the theoretical amplitudes in the Balmer line cores systematically deviate from those observed: while they are similar to, or slightly lower than the measurements for the higher Balmer lines (above H$\epsilon$), they increasingly overestimate the observed amplitudes in the lines at higher wavelengths (for H$\epsilon$, H$\delta$, and especially H$\gamma$ and H$\beta$). We are not sure whether this is due to some observational effect or caused by missing ingredients in our models (particularly concerning the chemical composition of the atmosphere). 
At the request of the referee we re-computed the $Q$-value of the fit taking into account only the continuum and the extended line wings (specifically, we masked out bands $\pm$6\AA wide centered around H$\beta$, HeII 4686, H$\gamma$, H$\delta$, H$\epsilon$ and H8). As expected, the quality-of-fit is improved, with $Q$=0.999 ($\ell$=0), $Q$=0.503 ($\ell$=1), $Q$=0.223 ($\ell$=2) and $Q\ll$0.001 for higher $\ell$ values. This is simply associated with the fact that there are now less deviations between the predicted and observed amplitudes, and nicely illustrates the very good match achieved for a radial mode in the continuum.

While we are conviced that the dominant mode in EC 20338$-$1925 is radial from the amplitude data alone, we also analysed the corresponding phase data as a consistency check. Given the high noise level of the measurements, we assumed a degree of $\ell=$ 0 from the outset, and attempted to fit the theoretical phases to those observed. In order to be able to compute the monochromatic phases accurately, we extended our existing code for the computation of the pulsational flux perturbation by an optional radial velocity (RV) parameter. This parameter was not necessary for the theoretical quantities presented by \citet{randall2005} because in that study we were interested only in wavelength-integrated phase measurements, which are dominated by the contributions from the continuum. As can be seen from Fig. \ref{ampphase}, these yield a relatively constant phase as a function of wavelength, which is consistent with the very small phase shifts predicted from one filter bandpass to the next in \citet{randall2005}. However, in a spectral line the moving matter produces significant distortions and phase delays over a pulsation cycle, and these need to be taken into account for monochromatic phase calculations. The distortions also affect the amplitudes in the spectral lines, but the effect is thought to be relatively small for the broader lines and was not included in the computations presented above for technical reasons. It remains to be investigated whether an incorporation of the Doppler shifts in the monochromatic amplitude computations could help improve the match between observations and theory in the Balmer lines.

The RV parameter incorporated into the phase calculations corresponds to the measurement of the semi-amplitude of the RV integrated over the disk, the same quantity that has been the subject of numerous time-series spectroscopy studies to date. Unfortunately, our data are simply not suitable for such an analysis due to the relatively low resolution, problems with the wavelength calibration (see Section 2), and spurious wavelength shifts caused by the star moving within the rather wide slit. We therefore have no observational constraint on the RV parameter, and instead treat it as a free parameter while fitting the observed monochromatic phases. That way, we were able to model the observed phases quite nicely, as can be seen in the right-hand panel of Fig. \ref{zoomampphase}. Note that the y-axis now measures the phase in radians rather than in seconds as in Fig. \ref{ampphase} for easier comparison to the computed values. This causes an apparent inversion of the curve due to a minus sign in the relation between phases specified in radians and those quoted in seconds. The radial velocity parameter used in the plot has a semi-amplitude of 11~km/s, which is quite similar to the values found for other high-amplitude EC 14026 pulsations from RV studies (e.g. $\sim$ 12 km/s for the dominant mode in PG 1325+101 from \citealt{telting2004}, 19~km/s for the unusually strong pulsation in Balloon 09010001 from \citealt{telting2006}, and 15~km/s for the main frequency in PG 1605+072 from \citealt{otoole2005}). We believe that this consolidates our interpretation of the dominant pulsation as a radial mode. Nevertheless, it is clear that given the observational limitations on the S/N as well as the lack of an independent radial velocity estimate, the discriminative power for the degree index of the mode lies in the amplitude, rather than the phase information.

Considering the lower amplitude modes extracted from our data, the monochromatic quantities are a lot more noisy, making a unique mode identification impossible. Nevertheless, we repeated the exercise of overplotting the normalised theoretical amplitude curves on the measurements for $f_2$ and $f_3$ (the data for $f_4$ were too noisy due to the much lower amplitude).  We used exactly the same theoretical values as for the dominant mode, because the small variation due to the difference in period is negligible when dealing with data this noisy. The results of the overplotting are shown in Fig. \ref{ampsf2f3}. Unsurprisingly, the S/N of the observations is too low to distinguish between the low-order degree modes with $\ell=$ 0,1 and 2, however the higher degree modes present an inferior fit, and can quite safely be excluded. The corresponding monochromatic phases meanwhile are too noisy to gain any information whatsoever.

\section{Conclusion}

In this study we attempted a mode identification of the main pulsation frequencies in the rapid sdB pulsator EC 20338$-$1925 on the basis of monochromatic amplitude and phase variations. We were able to obtain 3 half-nights of time-series spectrophotometry with the VLT instrument FORS2 in HIT-MS mode and detect four pulsations above the 4-$\sigma$ threshold. The S/N of the data was sufficient to extract high-quality monochromatic amplitudes and phases only for the dominant pulsation, noisier results being obtained for the remaining periodicities. Using full model atmosphere codes appropriate for the atmospheric parameters of our target and also incorporating non-adiabatic effects, we computed accurate theoretical amplitudes and phases for comparison to the observations. From the fits achieved to the observed amplitudes alone we determined the dominant pulsation to be a radial mode. This conclusion was found to be consistent with the monochromatic phase shifts observed under the assumption of a very reasonable radial velocity variation of 11~km/s. For two lower amplitude periodicities the data were found to be of sufficient quality to exclude higher degree modes with $\ell\geq$ 3.

To our knowledge, this is the first time that mode identification has been attempted for an EC 14026 star using monochromatic amplitudes. The fact that we were able to identify the main periodicity as a radial mode is therefore quite encouraging for future studies, particularly considering the relative faintness of our target. Moreover, the excellent match between theoretical predictions and the observations in the continuum confirms the basic validity of our approach and our models, although the discrepancies in the line cores indicate that there is still room for improvement, especially with respect to the chemical composition assumed in the model atmosphere calculations. Unfortunately, sdB stars are all chemically peculiar and contain heavy elements in varying abundances, implying that for a perfect match each target would first have to be submitted to a detailed abundance analysis and then analysed using a specially computed grid of model atmospheres. This is extremely time consuming both on the observational and the computational front, and not necessary for the practical application presented here, given that we have proved mode identification to be possible on the basis of the continuum behaviour alone.     

It is quite striking that in the few cases where observational mode identification has been possible for rapidly pulsating subdwarf B stars, the dominant mode of pulsation was always found to be a radial mode (for Balloon 090100001 from \citealt{charp2008}, KPD 2109+4401 from \citealt{randall2005}, HS 2201+2610 from Silvotti et al. 2010, submitted, and now for EC 20338-1925). Of course, this cannot automatically be assumed to hold true for all the highest amplitude frequencies detected in EC 14026 stars, but it does seem reasonable to deduce that mode visibility across the disk of the star is a very important factor influencing the observed amplitude. Asteroseismological solutions assigning high-degree modes to the strongest frequencies observed can in all likelyhood be disregarded from the outset, as has occasionally been done in the past to distinguish between several possible ``optimal models'' \citep[e.g.][]{randall2009}. This approach allows unambiguous model solutions to be found on the basis of fewer observed periods than if all observed frequencies are allowed to take on degree indices of $\ell$=0,1,2, and 4, and can therefore save valuable observing time.     

One puzzling implication of our identification of the 146.9-s periodicity as a radial mode is that this frequency cannot then contain closely spaced components due to a rotationally split multiplet. Consequently, it is difficult to explain the strong amplitude variations observed over several years (see Fig. 3) in terms of the beating of unresolved frequencies. Given that the photometry data obtained by Dave Kilkenny have time baselines of several days, the two independent harmonic frequencies would have to be spaced less than $\sim$ 2 $\mu$Hz apart in order to not be resolved. Additionally, the strong amplitude variations observed can be produced only if the two pulsations have comparable amplitudes, effectively implying two very closely spaced modes with degree indices $\ell$ = 0 and/or 1. From our stellar models we know that the required proximity in frequency is impossible for modes of the same degree index, and extremely unlikely for the combination of $\ell=$ 0 and $\ell=$ 1 \citep[see, e.g.][]{charp2000}. This is particularly true for a target with as high a surface gravity as EC 20338$-$1925, since the period density predicted is lower than for a more typical, less compact EC 14026 star. Therefore, we believe there is a high probability that the observed amplitude variations for the main mode are intrinsic to the star.

In conclusion, we have shown that the monochromatic amplitude and phase variations can be effectively used for mode identification in rapidly pulsating subdwarf B stars.  The most critical factor is the quality of the observational data, as the S/N level attained limits the discriminative power especially between modes of low degree indices with $\ell\leq$ 2. Even though we were allocated observing time using a highly suitable instrument mounted on one of the world's largest telescopes we were able to unambiguously identify the degree index for only the dominant, rather high amplitude mode. On the other hand, EC 20338$-$1925 is relatively faint, and there are several brighter targets available. Moreover, our observing run was affected by severe technical problems, due to which we lost a quarter of our allocated time completely, and did not obtain enough red data to constructively include it in our analysis. Therefore it seems likely that better results could be obtained in future similar studies.

It is not straightforward to objectively assess the relative merits of the different techniques currently in use for mode identification in subdwarf B stars, and opinions on this will vary. We believe that for the time being the exploitation of the pulsational amplitude's signature on $\ell$ as a function of wavelength is the most promising method, and it is so far the only one to have yielded unambiguous determinations of the degree index $\ell$. Given that the work presented here constitutes the first study employing monochromatic rather than broadband amplitudes, it is too early to say which of these two closely related techniques is more efficient. One drawback of the amplitude-wavelength approach to mode identification is that medium-sized or large telescopes coupled with specialised instruments are needed in order to obtain rapid time-series data of sufficient quality. However, this is equally true for the study of line-profile variations, where the low S/N achievable even with the world's largest telescopes has so far prevented unambiguous mode identification. Radial velocity surveys based on data from smaller telescopes have been successfully used to confirm periodicities previously detected from photometry, but we lack the theoretical framework necessary to use the information gained for mode identification. When  supplementing simultaneous multi-colour photometry, radial velocity measurements can indeed improve the discriminative power of the former, as was shown by \citet{baran2008,baran2010}. However, we would like to point out that for Balloon 09010001, the only case where a direct comparison of the two techniques is possible, four nights of standalone high S/N multi-colour photometry from a medium-size telescope \citep{charp2008} yielded an identification for significantly more modes than the combined results from 30 hours of simultaneous time-series spectroscopy obtained on a medium-size telescope and 120 hours of multi-colour photometry from a small telescope \citep{baran2008}. Therefore, we are confident that the exploitation of multi-wavelength amplitude information along the lines presented here makes the most efficient use of telescope time (at least until it becomes possible to obtain accurate monochromatic amplitudes and radial velocities with a single instrument),  and will prove highly conducive to mode identification and, consequently, asteroseismology in the future.

\begin{acknowledgements}
S.K.R. would like to thank the Paranal Science Operations, Engineering and Software staff, in particular Kieran O'Brien and Pascal Robert, for working hard to get FORS2 ready for these observations. We are grateful to Heidi Korhonen for her input regarding the wavelength calibration, and to Dave Kilkenny for providing us with an unpublished list of pulsation frequencies and amplitudes. G.F. also wishes to acknowledge the contribution of the Canada Research Chair Program.
\end{acknowledgements}

\bibliographystyle{aa}
\bibliography{ms_14780}

\begin{thebibliography}{43}
\expandafter\ifx\csname natexlab\endcsname\relax\def\natexlab#1{#1}\fi

\bibitem[{{Baran} {et~al.}(2008){Baran}, {Pigulski}, \& {O'Toole}}]{baran2008}
{Baran}, A., {Pigulski}, A., \& {O'Toole}, S.~J. 2008, \mnras, 385, 255

\bibitem[{{Baran} {et~al.}(2010){Baran}, {Telting}, {{\O}stensen}, {Winiarski},
  {Dro{\.z}d{\.z}}, {Kozie{\l}}, {Reed}, {Oreiro}, {Silvotti}, {Siwak},
  {Heber}, \& {Papics}}]{baran2010}
{Baran}, A., {Telting}, J., {{\O}stensen}, R., {et~al.} 2010, \apss, 32

\bibitem[{{Blanchette} {et~al.}(2008){Blanchette}, {Chayer}, {Wesemael},
  {Fontaine}, {Fontaine}, {Dupuis}, {Kruk}, \& {Green}}]{blanchette2008}
{Blanchette}, J., {Chayer}, P., {Wesemael}, F., {et~al.} 2008, \apj, 678, 1329

\bibitem[{{Brassard} {et~al.}(1995){Brassard}, {Fontaine}, \&
  {Wesemael}}]{brassard1995}
{Brassard}, P., {Fontaine}, G., \& {Wesemael}, F. 1995, \apjs, 96, 545

\bibitem[{{Brassard} {et~al.}(1992){Brassard}, {Pelletier}, {Fontaine}, \&
  {Wesemael}}]{brassard1992}
{Brassard}, P., {Pelletier}, C., {Fontaine}, G., \& {Wesemael}, F. 1992, \apjs,
  80, 725

\bibitem[{{Charpinet} {et~al.}(2009){Charpinet}, {Brassard}, {Fontaine},
  {Green}, {van Grootel}, {Randall}, \& {Chayer}}]{charp2009}
{Charpinet}, S., {Brassard}, P., {Fontaine}, G., {et~al.} 2009, in American
  Institute of Physics Conference Series, Vol. 1170, American Institute of
  Physics Conference Series, ed. {J.~A.~Guzik \& P.~A.~Bradley}, 585--596

\bibitem[{{Charpinet} {et~al.}(2001){Charpinet}, {Fontaine}, \&
  {Brassard}}]{charp2001}
{Charpinet}, S., {Fontaine}, G., \& {Brassard}, P. 2001, \pasp, 113, 775

\bibitem[{{Charpinet} {et~al.}(2005){Charpinet}, {Fontaine}, {Brassard},
  {Bill{\`e}res}, {Green}, \& {Chayer}}]{charp2005b}
{Charpinet}, S., {Fontaine}, G., {Brassard}, P., {et~al.} 2005, \aap, 443, 251

\bibitem[{{Charpinet} {et~al.}(1997){Charpinet}, {Fontaine}, {Brassard},
  {Chayer}, {Rogers}, {Iglesias}, \& {Dorman}}]{charp1997}
{Charpinet}, S., {Fontaine}, G., {Brassard}, P., {et~al.} 1997, \apjl, 483,
  L123+

\bibitem[{{Charpinet} {et~al.}(1996){Charpinet}, {Fontaine}, {Brassard}, \&
  {Dorman}}]{charp1996}
{Charpinet}, S., {Fontaine}, G., {Brassard}, P., \& {Dorman}, B. 1996, \apjl,
  471, L103+

\bibitem[{{Charpinet} {et~al.}(2000){Charpinet}, {Fontaine}, {Brassard}, \&
  {Dorman}}]{charp2000}
{Charpinet}, S., {Fontaine}, G., {Brassard}, P., \& {Dorman}, B. 2000, \apjs,
  131, 223

\bibitem[{{Charpinet} {et~al.}(2008{\natexlab{a}}){Charpinet}, {Fontaine},
  {Brassard}, {Randall}, {van Grootel}, {Green}, \& {Chayer}}]{charp2008b}
{Charpinet}, S., {Fontaine}, G., {Brassard}, P., {et~al.} 2008{\natexlab{a}},
  in Astronomical Society of the Pacific Conference Series, Vol. 392, Hot
  Subdwarf Stars and Related Objects, ed. {U.~Heber, C.~S.~Jeffery, \&
  R.~Napiwotzki}, 297--+

\bibitem[{{Charpinet} {et~al.}(2008{\natexlab{b}}){Charpinet}, {van Grootel},
  {Reese}, {Fontaine}, {Green}, {Brassard}, \& {Chayer}}]{charp2008}
{Charpinet}, S., {van Grootel}, V., {Reese}, D., {et~al.} 2008{\natexlab{b}},
  \aap, 489, 377

\bibitem[{{Clemens} {et~al.}(2000){Clemens}, {van Kerkwijk}, \&
  {Wu}}]{clemens2000}
{Clemens}, J.~C., {van Kerkwijk}, M.~H., \& {Wu}, Y. 2000, \mnras, 314, 220

\bibitem[{{Dorman} {et~al.}(1993){Dorman}, {Rood}, \& {O'Connell}}]{dorman1993}
{Dorman}, B., {Rood}, R.~T., \& {O'Connell}, R.~W. 1993, \apj, 419, 596

\bibitem[{{Fontaine} \& {Brassard}(1994)}]{fontaine1994}
{Fontaine}, G. \& {Brassard}, P. 1994, in Astronomical Society of the Pacific
  Conference Series, Vol.~57, Stellar and Circumstellar Astrophysics, a 70th
  birthday celebration for K. H. Bohm and E. Bohm-Vitense, ed. G.~{Wallerstein}
  \& A.~{Noriega-Crespo}, 195--+

\bibitem[{{Geier} {et~al.}(2007){Geier}, {Nesslinger}, {Heber}, {Przybilla},
  {Napiwotzki}, \& {Kudritzki}}]{geier2007}
{Geier}, S., {Nesslinger}, S., {Heber}, U., {et~al.} 2007, \aap, 464, 299

\bibitem[{{Han} {et~al.}(2003){Han}, {Podsiadlowski}, {Maxted}, \&
  {Marsh}}]{han2003}
{Han}, Z., {Podsiadlowski}, P., {Maxted}, P.~F.~L., \& {Marsh}, T.~R. 2003,
  \mnras, 341, 669

\bibitem[{{Han} {et~al.}(2002){Han}, {Podsiadlowski}, {Maxted}, {Marsh}, \&
  {Ivanova}}]{han2002}
{Han}, Z., {Podsiadlowski}, P., {Maxted}, P.~F.~L., {Marsh}, T.~R., \&
  {Ivanova}, N. 2002, \mnras, 336, 449

\bibitem[{{Heber}(2009)}]{heber2009}
{Heber}, U. 2009, \araa, 47, 211

\bibitem[{{Heber} {et~al.}(2000){Heber}, {Reid}, \& {Werner}}]{heber2000}
{Heber}, U., {Reid}, I.~N., \& {Werner}, K. 2000, \aap, 363, 198

\bibitem[{{Hubeny} \& {Lanz}(1995)}]{hubeny1995}
{Hubeny}, I. \& {Lanz}, T. 1995, \apj, 439, 875

\bibitem[{{Jeffery} {et~al.}(2005){Jeffery}, {Aerts}, {Dhillon}, {Marsh}, \&
  {G{\"a}nsicke}}]{jeffery2005}
{Jeffery}, C.~S., {Aerts}, C., {Dhillon}, V.~S., {Marsh}, T.~R., \&
  {G{\"a}nsicke}, B.~T. 2005, \mnras, 362, 66

\bibitem[{{Jeffery} {et~al.}(2004){Jeffery}, {Dhillon}, {Marsh}, \&
  {Ramachandran}}]{jeffery2004}
{Jeffery}, C.~S., {Dhillon}, V.~S., {Marsh}, T.~R., \& {Ramachandran}, B. 2004,
  \mnras, 352, 699

\bibitem[{{Jeffery} \& {Pollacco}(2000)}]{jeffery2000}
{Jeffery}, C.~S. \& {Pollacco}, D. 2000, \mnras, 318, 974

\bibitem[{{Kilkenny}(2010)}]{Kilkenny2010}
{Kilkenny}, D. 2010, \apss, 82

\bibitem[{{Kilkenny} {et~al.}(1997){Kilkenny}, {Koen}, {O'Donoghue}, \&
  {Stobie}}]{kilkenny1997}
{Kilkenny}, D., {Koen}, C., {O'Donoghue}, D., \& {Stobie}, R.~S. 1997, \mnras,
  285, 640

\bibitem[{{Kilkenny} {et~al.}(2006){Kilkenny}, {Stobie}, {O'Donoghue}, {Koen},
  {Hambly}, {MacGillivray}, \& {Lynas-Gray}}]{kilkenny2006}
{Kilkenny}, D., {Stobie}, R.~S., {O'Donoghue}, D., {et~al.} 2006, \mnras, 367,
  1603

\bibitem[{{Lanz} \& {Hubeny}(1995)}]{lanz1995}
{Lanz}, T. \& {Hubeny}, I. 1995, \apj, 439, 905

\bibitem[{{{\O}stensen} {et~al.}(2010){{\O}stensen}, {Oreiro}, {Solheim},
  {Heber}, {Silvotti}, {Gonz{\'a}lez-P{\'e}rez}, {Ulla}, {P{\'e}rez
  Hern{\'a}ndez}, {Rodr{\'{\i}}guez-L{\'o}pez}, \& {Telting}}]{ostensen2010}
{{\O}stensen}, R.~H., {Oreiro}, R., {Solheim}, J., {et~al.} 2010, \aap, 513,
  A6+

\bibitem[{{O'Toole} {et~al.}(2005){O'Toole}, {Heber}, {Jeffery}, {Dreizler},
  {Schuh}, {Woolf}, {Falter}, {Green}, {For}, {Hyde}, {Kjeldsen}, {Mauch}, \&
  {White}}]{otoole2005}
{O'Toole}, S.~J., {Heber}, U., {Jeffery}, C.~S., {et~al.} 2005, \aap, 440, 667

\bibitem[{{Press} {et~al.}(1986){Press}, {Flannery}, \&
  {Teukolsky}}]{press1986}
{Press}, W.~H., {Flannery}, B.~P., \& {Teukolsky}, S.~A. 1986, {Numerical
  recipes. The art of scientific computing}, ed. {Press, W.~H., Flannery,
  B.~P., \& Teukolsky, S.~A.}

\bibitem[{{Randall} {et~al.}(2005){Randall}, {Fontaine}, {Brassard}, \&
  {Bergeron}}]{randall2005}
{Randall}, S.~K., {Fontaine}, G., {Brassard}, P., \& {Bergeron}, P. 2005,
  \apjs, 161, 456

\bibitem[{{Randall} {et~al.}(2009){Randall}, {van Grootel}, {Fontaine},
  {Charpinet}, \& {Brassard}}]{randall2009}
{Randall}, S.~K., {van Grootel}, V., {Fontaine}, G., {Charpinet}, S., \&
  {Brassard}, P. 2009, \aap, 507, 911

\bibitem[{{Telting} {et~al.}(2010){Telting}, {{\O}stensen}, {Oreiro}, {Heber},
  {Vu{\v c}kovi{\'c}}, {Randall}, \& {Baran}}]{telting2010}
{Telting}, J., {{\O}stensen}, R., {Oreiro}, R., {et~al.} 2010, \apss, 88

\bibitem[{{Telting} {et~al.}(2008){Telting}, {Geier}, {{\O}stensen}, {Heber},
  {Glowienka}, {Nielsen}, {Oreiro}, \& {Frandsen}}]{telting2008}
{Telting}, J.~H., {Geier}, S., {{\O}stensen}, R.~H., {et~al.} 2008, \aap, 492,
  815

\bibitem[{{Telting} \& {{\O}stensen}(2004)}]{telting2004}
{Telting}, J.~H. \& {{\O}stensen}, R.~H. 2004, \aap, 419, 685

\bibitem[{{Telting} \& {{\O}stensen}(2006)}]{telting2006}
{Telting}, J.~H. \& {{\O}stensen}, R.~H. 2006, \aap, 450, 1149

\bibitem[{{Tremblay} {et~al.}(2006){Tremblay}, {Fontaine}, {Brassard},
  {Bergeron}, \& {Randall}}]{tremblay2006}
{Tremblay}, P.-E., {Fontaine}, G., {Brassard}, P., {Bergeron}, P., \&
  {Randall}, S.~K. 2006, \apjs, 165, 551

\bibitem[{{van Grootel} {et~al.}(2008){van Grootel}, {Charpinet}, {Fontaine},
  {Brassard}, {Green}, {Chayer}, \& {Randall}}]{val2008b}
{van Grootel}, V., {Charpinet}, S., {Fontaine}, G., {et~al.} 2008, \aap, 488,
  685

\bibitem[{{van Kerkwijk} {et~al.}(2000){van Kerkwijk}, {Clemens}, \&
  {Wu}}]{vankerkwijk2000}
{van Kerkwijk}, M.~H., {Clemens}, J.~C., \& {Wu}, Y. 2000, \mnras, 314, 209

\bibitem[{{Vu{\v c}kovi{\'c}} {et~al.}(2009){Vu{\v c}kovi{\'c}}, {{\O}stensen},
  {Aerts}, {Telting}, {Heber}, \& {Oreiro}}]{maja2009}
{Vu{\v c}kovi{\'c}}, M., {{\O}stensen}, R.~H., {Aerts}, C., {et~al.} 2009,
  \aap, 505, 239

\bibitem[{{Yu} \& {Li}(2009)}]{yu2009}
{Yu}, S. \& {Li}, L. 2009, \aap, 503, 151

\end{thebibliography}

\end{document}